# Intensity interferometry: Optical imaging with kilometer baselines

Dainis Dravins*

Lund Observatory, Box 43, SE-22100 Lund, Sweden

**ABSTRACT**

Optical imaging with microarcsecond resolution will reveal details across and outside stellar surfaces but requires kilometer-scale interferometers, challenging to realize either on the ground or in space. Intensity interferometry, electronically connecting independent telescopes, has a noise budget that relates to the electronic time resolution, circumventing issues of atmospheric turbulence. Extents up to a few km are becoming realistic with arrays of optical air Cherenkov telescopes (primarily erected for gamma-ray studies), enabling an optical equivalent of radio interferometer arrays. Pioneered by Hanbury Brown and Twiss, digital versions of the technique have now been demonstrated, reconstructing diffraction-limited images from laboratory measurements over hundreds of optical baselines. This review outlines the method from its beginnings, describes current experiments, and sketches prospects for future observations.

**Keywords:** Intensity interferometry, Cherenkov telescopes, Optical imaging, Long baselines, Stars, Stellar surfaces

## 1. HIGHEST ANGULAR RESOLUTION IN ASTRONOMY

Much of the progress in astronomy is driven by imaging with improved spatial resolution. Science cases are numerous in all branches of astronomy, from solar surface structure to sources of gravitational waves; from radio to gamma rays. The currently highest angular resolution realized in astronomical imaging is that by radio interferometers at their highest frequencies, operating between Earth and antennas in deep space[1-2]. Using the *RadioAstron* space antenna together with ground-based telescopes, a resolution of 21 μarcsec at 43 GHz has been demonstrated[3]. Through indirect techniques, indications of the presence of even much smaller structures can be found, even if explicit images are not obtained. Such 'superresolution' techniques include the use of turbulent scattering of radio waves in interstellar seeing[4] or the combination of the direct radio signal with that refracted or scattered off other cosmic structures, akin to classical sea interferometers[5-6]. Thus, an 'interstellar interferometer' may be formed, in special cases providing a spatial resolution on scales down to ~1 μas[7-9], or else aspects of small-scale source structure may be inferred from its time variability.

### 1.1 Highest angular resolution in optical astronomy

For other wavelength domains, the performance realized in radio remains a subject of envy. The current state of the art in the optical is represented by phase/amplitude interferometers such as the CHARA in California, MROI in New Mexico, NPOI in Arizona, SUSI in Australia, VLTI in Chile, and others. With baselines up to a few hundred meters and typically operating in the near-infrared, these realize resolutions on the order of 1 milliarcsecond (mas). Tantalizing results show how giant stars (diameters of a few tens of mas) are beginning to reveal themselves as objects with a broad variety of properties: flattened or deformed due to rapid rotation, engulfed by shells or obscuring clouds. However, typical bright stars in the sky subtend diameters of no more than a few mas and thus require baselines over at least several hundreds of meters or a few kilometers to enable any sensible surface imaging.

However, the stability requirements on amplitude (phase-) interferometers, together with the atmosphere above, to within a small fraction of an optical wavelength, constrain their operation for long baselines, especially at shorter visual wavelengths. The great scientific potential of stellar surface imaging has been realized by several, and concepts for long-baseline interferometers have been worked out for construction at ground-based observatories[10-11], in Antarctica[12], as free-flying telescopes in space[13-14], or even placed on the Moon[15-16]. Further concepts include the use of giant diffraction screens in space[17]. All these proposals are based upon sound physical and optical principles, yet do not appear likely to be realized in any immediate future due to their complexity and likely cost. Is there really no easier way?

*dainis@astro.lu.se, www.astro.lu.se/~dainis

## 2. INTERFEROMETRY IN OPTICAL ASTRONOMY

### 2.1 Optical amplitude interferometry: The beginnings

All current optical interferometers exploit the first-order interference of light waves, i.e., they relate the phases and amplitudes of light from a source that is being observed by two or more separated telescopes. The technique was originally proposed as an application of Young's double-slit experiment, first suggested by Fizeau[18], and then carried out on stars by Stéphan[19] in 1874. In the 1920's, Albert Michelson and co-workers developed an interferometer, mounted onto the then largest telescope, using a 6 meter baseline to measure the diameters of a few giant stars[20]. They then built a larger 15-m instrument[21-22] but found it too challenging to operate. The technique then lay dormant for half a century, until techniques had matured and Labeyrie[23] in 1975 succeeded in measuring interference fringes between two detached telescopes, triggering the construction of the current generation of optical interferometers.

### 2.2 Optical intensity interferometry: The beginnings

Intensity interferometry exploits a second-order effect of light waves, of the square of the light-wave amplitude, equivalent to its instantaneous intensity, and how the random (quantum) fluctuations in intensity correlate between light from a source that is being observed by two or more separated telescopes. The technique was developed by Robert Hanbury Brown and Richard Q. Twiss[24-25], who built an instrument with twin 6.5 m telescopes, movable along a circular railroad track at Narrabri in New South Wales, Australia.

The name 'intensity interferometer' is somewhat misleading since nothing actually is interfering; the name was selected for its analogy to the amplitude interferometer, which at that time had related scientific aims in measuring stellar diameters. It measures temporal correlations of arrival times between photons recorded in different telescopes. Initially, the understanding of its operation was a source of considerable confusion, and even now it may be challenging to intuitively comprehend. Seen in a quantum context, it builds upon a two-photon process, and is commonly seen as the first quantum-optical experiment. It laid the foundation for various experiments of photon correlations and for the development of the quantum theory of optical coherence, acknowledged with the 2005 Nobel prize in physics to Roy Glauber.

The great observational advantage of intensity interferometry (compared to amplitude interferometry) is that it is practically insensitive to either atmospheric turbulence or to telescopic optical imperfections, enabling very long baselines as well as observing at short optical wavelengths, even through large airmasses far away from the zenith. Telescopes are connected only electronically (rather than optically), from which it follows that the noise budget relates to the relatively long electronic timescales (nanoseconds, and light-travel distances of tens of centimeters or meters) rather than those of the light wave itself (femtoseconds and nanometers). A realistic time resolution of perhaps 10 ns corresponds to 3 m light-travel distance, and the control of atmospheric path-lengths and telescope imperfections then only needs to correspond to some reasonable fraction of those 3 m.

The Narrabri observatory was successfully used to measure the angular sizes of a number of bright and hot stars but, following its program, the technique has largely been dormant for half a century, as far as astronomy is concerned. However, intensity interferometry has been vigorously pursued in other fields, both for studying optical light in the laboratory, and in analyzing interactions in high-energy particle physics. For laboratory studies of scattered light, photon correlation spectroscopy can be considered as intensity interferometry in the temporal (rather than spatial) domain, and is a tool to measure the temporal coherence of light, and to deduce its spectral broadening[26-27].

In subatomic particle physics, the same quantum phenomena of intensity correlations apply to all bosons, i.e., particles which – like photons – have integer quantum spin, sharing the same Bose–Einstein quantum statistics (as opposed to fermions with their half-integer spin and Fermi-Dirac statistics). The method has found a wide application in particle physics, where it is most often called 'HBT-interferometry' (for Hanbury Brown-Twiss), although also the terms 'femtoscopy' or just 'Bose–Einstein correlations' are used[28].

## 3. INTENSITY INTERFEROMETRY: THE PRINCIPLES

The basic concept of an intensity interferometer is sketched in Figure 1. In its simplest form, an intensity interferometer consists of two telescopes (or more coarse 'flux collectors'), each with a photon detector feeding one channel of a signal processor for temporally cross correlating the signals from the two telescopes. Pairs of separate telescopes are simultaneously measuring the 'random' intensity fluctuations in the light from some particular source with the highest practical time resolution, typically in the range 1-10 nanoseconds. With the telescopes sufficiently close to one another, the fluctuations measured in both telescopes are more or less simultaneous, and thus correlated in time, but when moving them apart, the fluctuations gradually become decorrelated. How rapidly this occurs for increasing telescope separations gives a measure of the spatial coherence, and thus of the spatial properties of the source. This signal is a measure of the second-order spatial coherence, the square of that visibility which would be observed in any classical amplitude interferometer, and the spatial baselines for obtaining any given angular resolution are thus the same as would be required in ordinary interferometry.

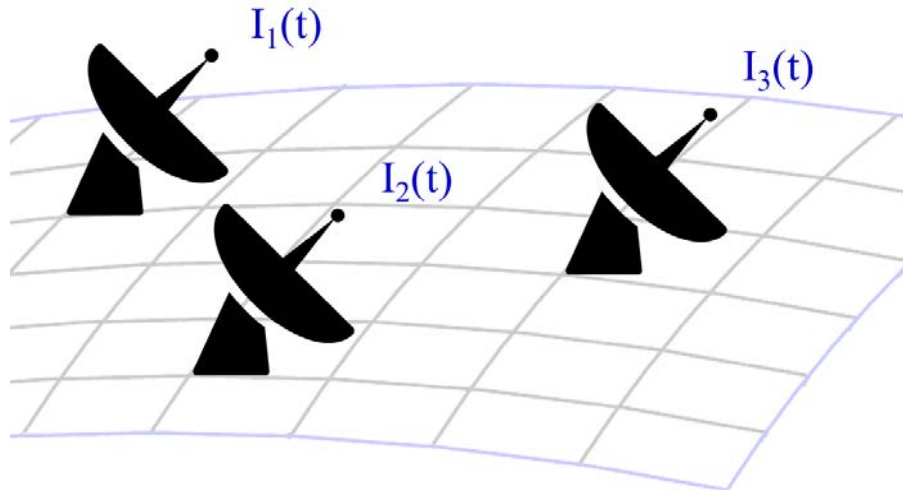

Figure 1. Components of an intensity interferometer. Spatially separated telescopes observe the same source, and the measured time-variable intensities $I_n(t)$ are electronically cross correlated between different pairs of telescopes. Two-telescope correlations are measured as $<I_1(t) \times I_2(t)>$, $<I_2(t) \times I_3(t)>$ and $<I_1(t) \times I_3(t)>$, three-telescope quantities as $<I_1(t) \times I_2(t) \times I_3(t)>$. There is no optical connection between the telescopes, and the operation resembles that of radio interferometers.

### 3.1 Intensity interferometry: Two-telescope observations

The quantity measured in intensity interferometry is: $<I_1(t) \times I_2(t)> = <I_1(t)><I_2(t)> (1 + |\gamma_{12}|^2)$, where $< >$ denotes temporal averaging and $\gamma_{12}$ is the mutual coherence function of light between locations 1 and 2, the quantity commonly measured in phase/amplitude interferometers (a relation that holds for each linear polarization). Compared to randomly fluctuating intensities, the correlation between intensities $I_1$ and $I_2$ is 'enhanced' by the coherence parameter and an intensity interferometer thus measures $|\gamma_{12}|^2$ with a certain electronic time resolution. Although this description is valid for a classical concept of light as a wave, this fundamentally is a quantum-optical two-photon effect, and presupposes that the light is in a state of thermodynamic equilibrium, obeying Bose-Einstein statistics. Such ordinary 'chaotic' light (also called 'thermal', 'maximum-entropy' or 'Gaussian') undergoes random phase jumps on timescales of its coherence time but the relations do not necessarily hold for light with different photon statistics (e.g., an ideal laser emits coherent light without any phase jumps, and thus would not generate any sensible signal in an intensity interferometer. Various monographs[29-31] provide more detailed discussions of the principal workings of intensity interferometry.

### 3.2 Intensity interferometry: Three or more telescopes

Greater differences between intensity and amplitude interferometry develop in the case of arrays with many telescopes. Since the intensity signal is electronic only, it may be freely copied, transmitted, combined or saved, quite analogous to radio interferometry. In any larger array, the possible number of baselines between telescope pairs grows rapidly, without any additional logistic effort. With $N$ telescopes, $N(N-1)/2$ baselines can be formed, even if possibly periodic telescope locations could make some of them redundant. With telescopes fixed on the ground, the projected baselines change during an observing night, as a source moves across the sky. The signal handling has then to assign successive measures of the spatial coherence $|\gamma_{12}|^2$ to the appropriate baseline length and orientation.

As mentioned above, the primary measures are cross correlations between intensity fluctuations at two spatial locations. However, one may construct also higher-order quantities, for example three-point intensity correlations for systems of three telescopes[32-34]: $<I_1(t) \times I_2(t) \times I_3(t)> = <I_1(t)><I_2(t)><I_3(t)> (1 + |\gamma_{12}|^2 + |\gamma_{23}|^2 + |\gamma_{31}|^2 + 2 \text{Re}[\gamma_{12}\gamma_{23}\gamma_{31}])$.

The phase of the triple product in the last term is the 'closure phase', a quantity widely used in amplitude interferometry to eliminate effects of differential atmospheric phase errors between telescopes since the baselines 1-2, 2-3, and 3-1 form a closed loop. Of course, intensity interferometry is not sensitive to phase errors, but three-point intensity correlations permit to obtain the real (cosine) part of this closure-phase function. Such third-order measures are more sensitive to observational noise but, on the other hand, large telescope arrays permit many different three-telescope combinations to be electronically created. Such three-point correlations provide additional constraints that may enhance image reconstruction[33-34]. Also, fourth- and higher-order correlations may be constructed from quadruplets or larger groups of telescopes, the effort, in principle, being solely in software. Such higher-order correlations in light may well carry additional information about the source[32,35-41].

## 4. REVIVAL OF INTENSITY INTERFEROMETRY IN ASTRONOMY

After the original Narrabri intensity interferometer, the technique has now largely been dormant for decades, as far as astronomy is concerned. A number of concurrent developments now make its return very timely. Over past decades, very significant progress and tantalizing discoveries have been made with optical amplitude interferometry but their stringent stability requirements hold back the construction of complexes comprising telescopes spread over kilometric baselines, as required for microarcsecond imaging. A further constraint for larger amplitude interferometry arrays is that optical light cannot be copied with retained phase but has to be split up (and thus attenuated) by multiple beamsplitters to achieve interference among numerous telescope pairs.

While intensity interferometry circumvents atmospheric turbulence and is immune to telescopic imperfections, it is so at a certain cost. Realistically, the coherence parameter $|\gamma_{12}|^2$ is measured with a time resolution of nanoseconds, which is much longer than the typical coherence time of broad-band light, over which the intensity fluctuations are fully developed. While, on one hand, such a time resolution relaxes the atmospheric and telescopic error budgets, it also averages out the observable intensity fluctuations over very many coherence times, so that large photon fluxes are required to reduce the statistical noise and enable precise measurements of a small signal. For this reason, astronomical intensity interferometry cannot be carried out with small telescopes (as opposed to amplitude interferometers); already the 6.5 m flux collectors in the original intensity interferometer at Narrabri were larger than any other optical telescope at that time (and even so, typical integration times on bright stars were on the order of several hours).

The most essential component for a large intensity interferometer is thus a group of large telescopes or flux collectors spread over some square kilometer or so. Just such complexes are being erected for a quite different primary purpose: air Cherenkov telescopes for the study of gamma-ray sources through the observation of the visual flashes of Cherenkov light emitted from the particle cascades initiated by the gamma rays in the upper atmosphere. Since these flashes are faint, the telescopes must be large (currently up to 28 m aperture) but do not need to be optically more precise than the spatial equivalent of a few nanoseconds light-travel time, that being the typical duration of these Cherenkov flashes. For good stereoscopic source localization, the telescopes need to be spread out over hundreds of meters, that being the typical extent of the light pool on the ground. These parameters are remarkably similar to the requirements for intensity interferometry and several authors have realized the potential also for this application[42-44]. In the original Narrabri instrument, the telescopes were continuously moved during observations to maintain their projected baseline. However, electronic time delays can now be used instead to compensate for different arrival times of a wavefront to the different telescopes, removing the need for having them mechanically mobile.

The largest current such project of air Cherenkov telescopes is CTA, the Cherenkov Telescope Array[45-46]. With two geographical sites, it is foreseen to ultimately have up to some 100 telescopes spread over several square kilometers, with a combined light-collecting area on the order of 10,000 m$^2$. Of course, it will mainly be devoted to its main task of observing Cherenkov light in air; however several other applications have been envisioned, preferably to be carried out during nights with bright moonlight which − due to the faintness of the Cherenkov light flashes − might preclude their efficient observation. Suggested additional uses besides intensity interferometry[47] include searches for rapid astrophysical events[48-50], observing stellar occultations by distant Kuiper-belt objects[51] or as a terrestrial ground station for optical communication with distant spacecraft[52-53]. If baselines of 2 or 3 km could be utilized with CTA at short optical wavelengths, resolutions would approach 30 μas, an unprecedented spatial resolution in optical astronomy[44-47].

# 5. APERTURE-SYNTHESIS IMAGING WITH LARGE OPTICAL ARRAYS

While a two-telescope interferometer (such as the classical one at Narrabri) is useful in determining the angular scales of sources, it offers only very limited coverage of the interferometric Fourier-transform (*u,v*) plane, and does not easily lend itself to two-dimensional imaging. Analogous to radio interferometers, that requires a multi-telescope grid to provide numerous baselines. However, as compared to amplitude interferometers, full imaging carries an additional challenge in that, while intensity data do provide the absolute magnitudes of the respective Fourier transform components of the source image, their phases are not directly obtained. On one hand, such Fourier magnitudes could well be used by themselves to fit model parameters such as stellar diameters, stellar limb darkening, binary separations, circumstellar disk thicknesses, etc., but actual images cannot be *directly* obtained through just a simple inverse Fourier transform.

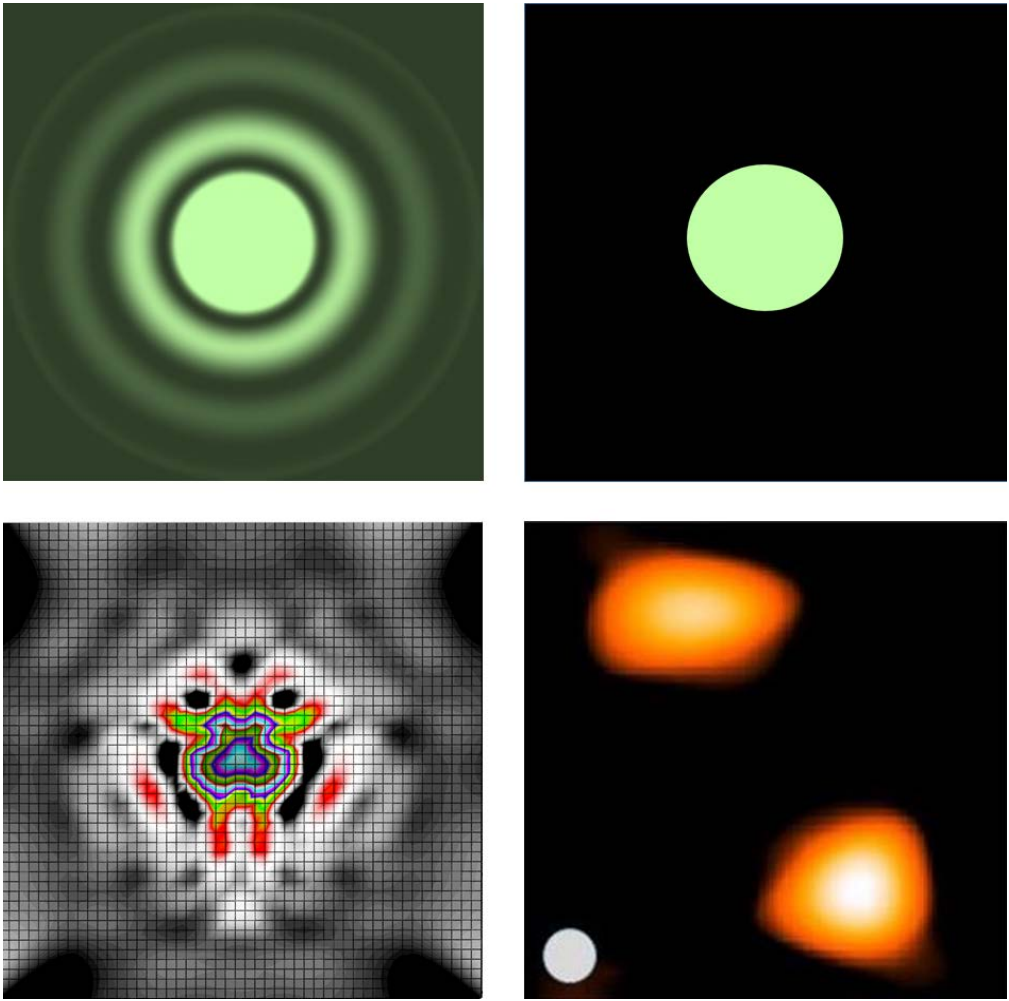

Figure 2. Fourier-plane and image-plane information. The left column shows two-dimensional coherence maps ('diffraction patterns') corresponding to the images at right. At top left, the familiar monochromatic Airy diffraction pattern can readily be recognized as originating from a circular aperture at right. At bottom left is an actually measured second-order optical coherence pattern (in false colors) from an artificial asymmetric binary star, built up from intensity-correlation measurements over 180 baselines between pairs of small laboratory telescopes. At right is the image reconstructed from these intensity-interferometric measurements[54-55]. The circle marks the diffraction-limited spatial resolution, thus realized by an array of optical telescopes connected through electronic software only, with no optical links between them.

Nevertheless, it is possible to recover the phase of a complex function when only its magnitude is known, and various techniques have been developed towards this aim (most unrelated to astronomy, e.g., for coherent diffraction imaging in X-rays). Already intuitively, it is clear that the information contained in the coherence map (equivalent to the source's diffraction pattern) must place stringent constraints on the source image. For instance, viewing the familiar Airy diffraction pattern in Figure 2, one can immediately recognize it as originating in a circular aperture, although only intensities are seen. However, it is also obvious that a reasonably complete coverage of the diffraction image is required to convincingly identify a circular aperture as the source. Imaging methods specifically for intensity interferometry have been worked out for one[56] and two dimensions[57-58]. Such phase recovery techniques have been applied to reconstruct images from simulated intensity interferometry observations[59-62], demonstrating that also rather complex images can be reconstructed. One remaining limitation is the non-uniqueness between the image and its mirrored reflection. Also that ambiguity, however, might be resolved by supplementing the data with, e.g., lower-resolution measurements from single telescopes, and enforcing image continuity. Figure 2 shows an example of actual measurements of an artificial star with an array of small optical telescopes in the laboratory, operated as an intensity interferometer with 180 baselines[54-55].

## 5.1 Signal-to-noise ratios in intensity interferometry

The observing program of the original intensity interferometer at Narrabri covered bright (down to second-magnitude) and relatively hot stars (~9000 K), while estimates of limiting magnitude for modern multi-telescope arrays are around visual magnitude 8. The signal-to-noise ratio depends primarily on a number of readily understandable parameters such as telescope sizes, detector efficiency, time resolution and total observing time[24,44,47,63].

A less obvious dependence is that on the intrinsic brightness temperature of the source, a property specific to intensity interferometry. For a given photon flux, the signal-to-noise ratio is better for sources where those photons are squeezed into narrower optical passbands. For a flat-spectrum source, the S/N actually is independent of the width of the optical passband, whether measuring only the limited light inside a narrow spectral feature or a much greater broad-band flux. This property was exploited already in the original Narrabri interferometer[64] to identify the extended emission-line volume from the stellar wind around the Wolf–Rayet star $\gamma^2$ Vel. The same effect could also be exploited for increasing the signal-to-noise by observing the same source simultaneously in multiple spectral channels, a concept foreseen for the once proposed successor to the original Narrabri interferometer[25].

This independence from optical passband of course cannot be extrapolated without limit, but it holds for all realistic electronic resolutions (~ns), which (almost) always are much slower than the temporal coherence time of light from astronomical sources. (Although, with elaborate filtering, extremely narrow passbands can be produced where these quantum fluctuations begin to get resolved[65-66].) While narrowing the spectral passband does decrease the photon count rate, it also increases the temporal coherence by the same factor. The intensity fluctuations have their full amplitude over one coherence time, and now get averaged over fewer coherence periods, canceling the effects of increased photon noise.

The practically achievable signal-to-noise ratios are being examined by different groups, who have been testing and simulating detectors, signal and data handling and observing procedures in laboratories, as well as at Cherenkov and conventional optical telescopes[67-78]. Thousands of sources in the sky are bright and hot enough to be mapped already with current instrumentation performance. In a quest to reach also much fainter sources, one major breakthrough can be expected once energy-resolving detectors become available to use for also high photon count rates. This would enable parallel observing in multiple spectral channels and increase the signal by a factor equal to the spectral resolution. Energy-resolving detectors already used in optical astronomy include superconducting tunnel junctions, transition-edge sensors, and microwave kinetic inductance detectors. However, these do not yet appear to have reached a level of development where routine operation at Cherenkov telescopes would be practical.

## 6. MICROARCSECOND ASTROPYSICS

Probably the 'easiest' targets for intensity interferometry observations are relatively bright and hot, single or binary stars of spectral types O and B or Wolf-Rayet stars with their various circumstellar emission-line structures. Their stellar disk diameters are typically ~ 0.2–0.5 mas and thus lie somewhat beyond what can be resolved with existing amplitude interferometers. Also rapidly rotating stars, with oblate shapes deformed by rotation, circumstellar disks, winds from hot stars, blue supergiants and extreme objects such as η Carinae, interacting binaries, the hotter parts of [super]nova explosions, pulsating Cepheids or other hotter variables are clear candidates[44,47,79]. Obviously, almost any source would benefit from higher resolution; the limit is set only by how faint the sources are that practically can be observed.

## 6.1 The true meaning of microarcsecond resolution

The full Cherenkov Telescope Array will have numerous telescopes distributed over a few square km, with an edge-to-edge distance of about two or three km. If fully equipped for intensity interferometry at the shortest optical wavelengths, the spatial resolution may approach ∼30 µas. Such resolutions have hitherto been reached only in the radio, and it is awkward to speculate upon which features could appear in optical sources. However, to appreciate the meaning of such resolutions, Figure 3 shows an 'understandable' type of object: known solar-type phenomena projected onto the (unknown) disk of a nearby star. A hypothetical exoplanet is shown in transit across the disk of Sirius. The planet's size and oblateness were taken equal to that of Jupiter, but fitted with a Saturn-like ring and four 'Galilean' moons, each the size of Earth. While spatially resolving the disk of an exoplanet in its reflected light may remain unrealistic, the imaging of its dark silhouette on a stellar disk – while certainly still very challenging – could perhaps be not quite impossible[80-82].

Intensity interferometry actually possesses some advantages for such possible observations. The lack of sensitivity to the phases of the Fourier components of the image could be an advantage since one would all the time measure 'only' the amplitude of the Fourier transform of the exoplanet image, irrespective of where on the stellar disk it happened to be. Unless close to the stellar limb, the star only serves as a bright background, its comparatively 'huge' diameter of 6 mas not contributing any sensible spatial power at any relevant telescopic baselines.

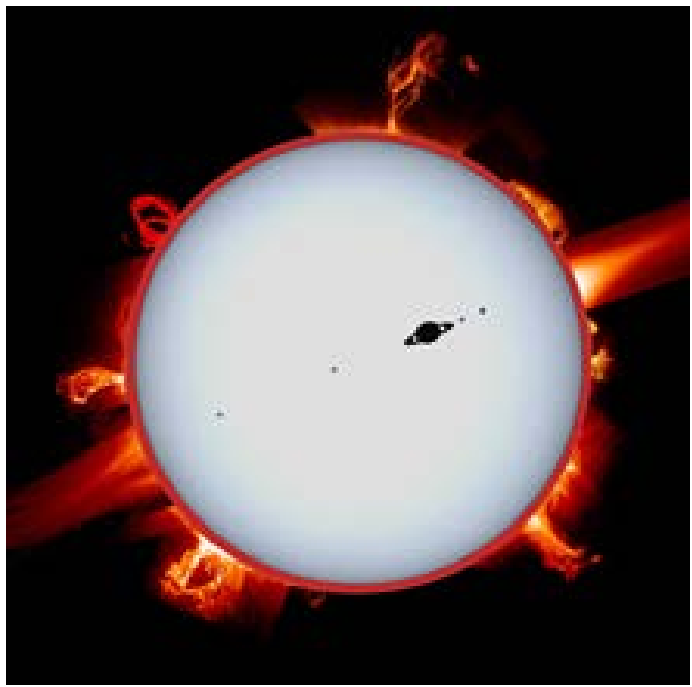

Figure 3. The real meaning of 40 microarcsecond optical resolution: Simulated resolution for an assumed transit of a hypothetical exoplanet across the disk of the relatively nearby star Sirius, using the full Cherenkov Telescope Array as an intensity interferometer. Stellar angular diameter = 6 mas; assumed planet of Jupiter size and oblateness; equatorial diameter = 350 µas; Saturn-type rings; four Earth-size moons. The stellar surface is assumed surrounded by a solar-type chromosphere, shining in an emission line. The 40 µas resolution provides some 150 pixels across this stellar diameter.

The photospheric stellar surface in Figure 3 is assumed to be surrounded by a solar-type chromosphere, shining in a reddish emission-line such as H-alpha, with various prominences and eruptions protruding. Separate observations of the white-light continuum, and the emission-line chromosphere are in principle straightforward since the S/N ratio is independent of optical bandpass and the brightness temperature in optically thick chromospheric lines may well be at least as high as that of the white-light photosphere.

Figure 3 is not a simulation of specific observations but just an image degraded to the angular resolution expected to be reachable with a complex such as the CTA. Certainly, a lot of work will have to be invested before such images actually can be retrieved, but there seem to be no fundamental obstacles to entering the domain of microarcsecond imaging.

# 7. BEYOND INTENSITY INTERFEROMETRY: ASTRONOMICAL QUANTUM OPTICS

The quantity measured in intensity interferometry is only one particular aspect of the second-order coherence of light, namely its value evaluated at two different spatial locations $r_1$ and $r_2$, but at one instant in time along the propagating wavefront, i.e., $g^{(2)}(r_{12},t_{12}) = <I_1(r_1,t_1) \times I_2(r_2,t_2)> / <I_1(r_1,t_1)><I_2(r_2,t_2)>$ for the special case $t_1 = t_2$. As already mentioned, the functioning of intensity interferometry presupposes that the distribution of photons in the light is 'thermal' or 'chaotic', rather than being, e.g., second-order coherent such as from an ideal laser. A hypothetical star shining with ideal laser light would not show any intensity fluctuations on any spatial baselines, nor over any temporal delays and therefore would give the impression of a white-light source of unlimited spatial extent. Such an example shows that there is more information content in light than can be recorded by an intensity interferometer (indeed more than can be seen by any classical astronomical instrument): we are here entering the realm of quantum optics of photon statistics, photon orbital angular momentum, and other properties of three-dimensional volumes of photon gases.

Besides parameters related to the first-order spatial and temporal coherence of light (manifest in ordinary spatial images and spectra), light has additional degrees of freedom, discernable in the statistics of photon arrival times, or in the amount of photon orbital angular momentum. Such quantum-optical measures may carry information on how the light was created at the source, and whether it reached the observer directly or via some intermediate process. Astronomical quantum optics may help to clarify emission processes in natural laser sources and in the environments of compact and energetic objects. Time resolutions of nanoseconds are required, as are large photon fluxes, making photonic astronomy very timely in an era of forthcoming extremely large telescopes[83-84].

For a stable wave (such as an idealized laser), $g^{(2)}(\tau) = g^{(2)}(0) = 1$ for all time delays; while for chaotic, maximum-entropy light used in intensity interferometry $g^{(2)}(0) = 2$ (if measured with perfect time resolution), a value reflecting the degree of photon bunching in the Bose-Einstein statistics of thermodynamic equilibrium. In the laboratory, one can follow how the physical nature of the photon gas gradually changes from chaotic [$g^{(2)} = 2$] to ordered [$g^{(2)} = 1$] when a laser is 'turned on' and the emission gradually changes from spontaneous to stimulated (although the spectrum does not change). Measuring $g^{(2)}$ and knowing the laser parameters involved, it is possible to deduce the atomic energy-level populations, which is an example of a parameter of significance to theoretical astrophysics ('non-LTE departure coefficient') which cannot be directly observed with any classical measurements of one-photon properties. Chaotic light that has been scattered by a Gaussian frequency-redistributing medium reaches a higher degree of photon bunching: $g^{(2)}(0) = 4$; while fully anti-bunched light has $g^{(2)}(0) = 0$. The latter state implies that whenever there is one photon at some particular time $= t$ (then $I(t) = 1$), there is none immediately afterwards, i.e., $I(t+\tau) = 0$ for sufficiently small $\tau$. Experimentally, this can be produced in, e.g., resonance fluorescence, and is seen through sub-Poissonian statistics of recorded photon counts, i.e. narrower distributions than would be expected in a 'random' situation. Corresponding relations exist for higher-order correlations, measuring the properties for groups of three, four, or a greater number of photons. Experimental procedures for studying such effects in non-astronomical contexts are described in various monographs[26-27,85].

Among astronomical applications, one can ask what is the quantum nature of the light emitted from a volume with departures from thermodynamic equilibrium of the atomic energy level populations? Will a spontaneously emitted photon stimulate others, so that the path where the photon train has passed becomes temporarily de-excited and remains so for perhaps a microsecond until collisions and other effects have restored the balance? Such temporal intensity correlations are observed in the laboratory when light is scattered by hot atomic vapor[86]. Does then light in a spectral line perhaps consist of short photon showers with one spontaneously emitted photon leading a trail of others emitted by stimulated emission? Such amplified spontaneous emission ('partial laser action') might occur in atomic emission lines from extended stellar envelopes or stellar active regions. Predicted locations are mass-losing high-temperature stars, where the rapidly recombining plasma in the stellar envelope can act as an amplifying medium[87-90]. By making astronomical observations through temporal correlations in software, effective spectral resolutions that are orders of magnitude beyond what is feasible with mechanical spectrometers could be achieved, possibly resolving the predicted narrow linewidths of cosmic lasers[91]. Further possible astrophysical effects detectable through photon statistics include collective spontaneous emission ('Dicke superradiance'). There are further challenges in quantum imaging beyond mere intensity interferometry: a somewhat related technique is 'ghost imaging'. Here, one exploits the presence of correlated photon pairs in natural light, where the scattering of light against also distant objects can act as a beamsplitter, possibly yielding detectable variations in the intensity-correlation signatures[92].

# 8. CONCLUSIONS

Long after the pioneering experiments by Hanbury Brown and Twiss, we are seeing great progress in electronic and computing technologies; mathematical algorithms have been developed for image reconstruction from intensity-correlation data and the most essential components for intensity interferometry – large and well separated light collectors – are being realized in the form of arrays of air Cherenkov telescopes. All this promises to achieve an unpretentious but difficult goal: to finally be able to view our neighboring stars not only as mere unresolved points of light but as the extended and most probably very fascinating and diverse objects that they really are.

# ACKNOWLEDGMENTS

This work has been supported by the Swedish Research Council and The Royal Physiographic Society in Lund. The development of concepts for intensity interferometry with Cherenkov telescope arrays has involved interactions with several colleagues elsewhere, in particular at the University of Utah in Salt Lake City (David Kieda, Stephan LeBohec, and Paul D. Nuñez) and at the University of Padova (Cesare Barbieri and Giampiero Naletto). Experiments towards laboratory intensity interferometry at Lund Observatory have involved also Toktam Calvén Aghajani, Colin Carlile, Hannes Jensen, Tiphaine Lagadec, Ricky Nilsson and Helena Uthas while laboratory studies of photon-counting detectors were made also by Daniel Faria and Johan Ingjald. I thank Colin Carlile for constructive comments on the manuscript.